\documentclass[amsmath,twocolumn,prl,aps]{revtex4}
\usepackage{amsthm,amsfonts,graphicx,verbatim}
\newcommand{\be}{\begin{equation}}
\newcommand{\ee}{\end{equation}}
\newcommand{\bea}{\begin{eqnarray}}
\newcommand{\eea}{\end{eqnarray}}

\newcommand{\lp}{\left(}
\newcommand{\rp}{\right)}

\renewcommand{\phi}{\varphi}
\renewcommand{\epsilon}{\varepsilon}
\renewcommand{\vec}[1]{{\bf #1}}

\begin{document}

\title{
Conformal Invariance and Shape-Dependent Conductance of Graphene Samples}
\author{
Dmitry A. Abanin$^1$, Leonid S. Levitov$^{1,2}$}
\affiliation{
 ${}^1$ Department of Physics,
 Massachusetts Institute of Technology, 77 Massachusetts Ave,
 Cambridge, MA 02139\\
${}^2$ Kavli Institute for Theoretical Physics, University of California, Santa Barbara, CA 93106
}
\date{\today}

\begin{abstract}
For a sample of an arbitrary shape, the dependence of its conductance on the longitudinal and Hall conductivity is identical to that of a rectangle.
We use analytic results for a conducting rectangle, combined with the semicircle model for transport coefficients, to study properties of the monolayer and bilayer graphene. 
A conductance plateau centered at the neutrality point, predicted for square geometry, is in agreement with recent experiments. For rectangular geometry, the conductance exhibits maxima at the densities of compressible quantum Hall states for wide samples, and minima for narrow samples. The positions and relative sizes of these features are different in the monolayer and bilayer cases, indicating that the conductance can be used as a tool for sample diagnostic. 
\end{abstract}

\maketitle
\section{Introduction}
After the observation of the quantized Hall effect (QHE) in graphene~\cite{Novoselov05,Zhang05}, this material has quickly moved into the focus of research on quantum transport. Recent advances in patterning graphene into quantum dots \cite{Geim07}, ribbons \cite{Chen07,Han07} and other shapes, and in selective gating of graphene sheets \cite{Huard07,Williams07,Ozyilmaz07}, has created a remarkably active field of graphene nanoelectronics. 
One of the challenges in these experiments is finding reliable methods for sample characterization, allowing to distinguish graphene monolayer from graphene bilayer or multilayer systems. Quite often, the means for that are provided by Raman spectroscopy~\cite{Ferrari06}.
However, in a number of experiments it is more convenient to perform sample characterization using transport measurements~\cite{private_comm}.

The simplest transport characteristic of a graphene device is its two-terminal conductance. This quantity exhibits plateau-like structure in the QHE regime, occurring at different electron densities in the monolayer and bilayer systems, which in principle makes it suitable for sample diagnostic. However, the observed conductance plateaus are often distorted, which
is not surprising, because the two-terminal conductance, unlike the resistivity obtained from a four-probe measurement, in general depends on the sample aspect ratio and other geometric characteristics. This dependence must be taken into account, in as much as possible, in interpreting the measurement results.

The effect of sample shape on the conductance can be illustrated by the well known formula
for a conducting square with the longitudinal and Hall conductivities $\sigma_{xx}$ and $\sigma_{xy}$ and ideal contacts on opposite sides~\cite{Lippmann58,Jensen72}:
\be\label{g_square}
G_{L=W}=\sqrt{\sigma_{xx}^2+\sigma_{xy}^2}
,
\ee
which follows from a duality relation for 2d transport (see ~\cite{Jensen72,Dykhne1970} and discussion below).
The result \eqref{g_square} gives the macroscopic conductance in terms of microscopic transport coefficients. As we shall see, the dependence on $\sigma_{xx}$ and $\sigma_{xy}$ in Eq.\eqref{g_square} is such that it can make the conductance $G_{L=W}$ density-independent in the QHE regime near the graphene charge neutrality point (CNP), where both $\sigma_{xx}$ and $\sigma_{xy}$ have strong density dependence. Thus the effect of sample geometry on conductance may be nonintuitive and should be accounted for carefully.

Is it possible to extend the result (\ref{g_square}), valid  for a perfect square, to other sample geometries? The next simplest shape to a square is a rectangle, for which the two-terminal conductance was found in Ref.\cite{Lippmann58} using conformal mappings of the Schwartz-Christoffel form. For all other shapes, luckily, the conduction problem can be reduced, via a conformal mapping, to that of a rectangle (see Appendix). The aspect ratio of such an equivalent rectangle, which depends on the size of the contacts and on their separation, can serve as a parameter that classifies all conduction problems.

In Ref.\cite{Lippmann58} a closed-form expression for the conductance of a rectangle was obtained via an integral representation. However, as discussed below, the integrals of Ref.\cite{Lippmann58} are convergent very slowly, especially in the interesting limit of large Hall angles. Because of that,  
for our purpose it will be more convenient to 
employ the approach  of Rendell and Girvin~\cite{Girvin81}, which describes spatial distribution of the electric field and current in terms of a suitably chosen analytic function, allowing for direct numerical evaluation of the conductance.

In this article, we use the method of Ref.~\cite{Girvin81} combined with the  effective medium approach of Dykhne and Ruzin~\cite{semicircle} that yields a semicircle relation between $\sigma_{xx}$ and $\sigma_{xy}$. We analyze conductance as a function of carrier density, focusing on the features that distinguish between transport in the monolayer and bilayer graphene. We conclude that the dependence of conductance on the sample shape, which may be quite strong, does not mask the difference between the monolayer and bilayer systems even in the absence of clear conductance plateaus.

For the effective medium approach \cite{semicircle} to be applicable, the sample size must be large compared to the typical charge inhomogeneity length scale $\xi$, otherwise strong mesoscopic sample-to-sample fluctuations are to be expected. In most of the paper we focus on the case of large samples, which can be described by a spatially uniform conductivity tensor obeying the semicircle relation. 
We shall briefly discuss the situation in mesoscopic samples of size comparable to $\xi$ at the end of the paper.

\section{Duality relation for conductance}

Here we shall focus on the rectangular geometry illustrated in Fig.\ref{fig2} inset (later, in Appendix, it will be shown that for any conductor shape the problem can be mapped on that of an equivalent rectangle). To describe electric transport, we employ the bulk conduction approach, in which the sample bulk is characterized by the longitudinal and Hall conductivities $\sigma_{xx}$, $\sigma_{xy}$. The transport equation is $\vec j=\hat\sigma\vec E$ where $\hat\sigma$ is a $2\times2$ conductivity tensor, with the current and electric field obeying
\be\label{div-curl}
\nabla\cdot \vec j = 0
,\quad
\nabla\times \vec E = 0
\ee
These equations must be solved with the boundary conditions $\vec j_\perp=0$ at $x=0,W$ (current continuity) and $\vec E_\parallel = 0$ at $y=\pm L/2$ (ideal ohmic contacts). 

It is instructive to apply duality transformation~\cite{Dykhne1970,Jensen72} to this problem, rotating current $\vec j$ and electric field $\vec E$ by $90^o$ and interchanging them: $\vec j'=R_{\pi/2}\vec E$, $\vec E'=R_{\pi/2}\vec j$. Upon such a transformation the transport equations (\ref{div-curl}), as well as the boundary conditions, preserve their form, whereby the conductivity tensor is replaced by $\hat \sigma'=\hat \sigma^{-1}$ and
the dimensions of the rotated rectangle interchange: $L'= W$, $W'= L$. Since resistance for the transformed problem $R=V'/I'=1/G'$ is identical to  the conductance of the initial problem $G=I/V$, where $V$ is source/drain voltage and $I$ is net current, we obtain a duality relation 
\be\label{duality0}
G(L,W,\hat \sigma)=G^{-1}(L',W',\hat \sigma')
\ee
We note that, since $\hat \sigma'=\hat \sigma^{-1}$ is the resistivity tensor,
the quantity $G(L',W',\hat \sigma')$ has dimension of resistivity, and so the right hand side of (\ref{duality0}) has dimension of conductance.
To simplify the relation (\ref{duality0}), we take into account that $G$ scales with $\hat \sigma$, i.e. that $G(L,W,\eta\hat \sigma)=\eta G(L,W,\hat \sigma)$,
and that it is invariant 
upon sign reversal of $\sigma_{xy}$. Writing $\sigma'_{xx}=\sigma_{xx}/(\sigma_{xx}^2+\sigma_{xy}^2)$, $\sigma'_{xy}=\sigma_{xy}/(\sigma_{xx}^2+\sigma_{xy}^2)$, and using the scaling property of $G$, we obtain a relation
\be\label{duality}
G(L,W,\hat \sigma)=(\sigma_{xx}^2+\sigma_{xy}^2)/G(W,L,\hat \sigma)
,
\ee
which connects the rectangles $L\times W$ and $W\times L$ having {\it the same} bulk transport coefficients. Setting $L=W$, 
we obtain the conductance of a square, Eq.(\ref{g_square}).

\begin{figure}
\includegraphics[width=3.4in]{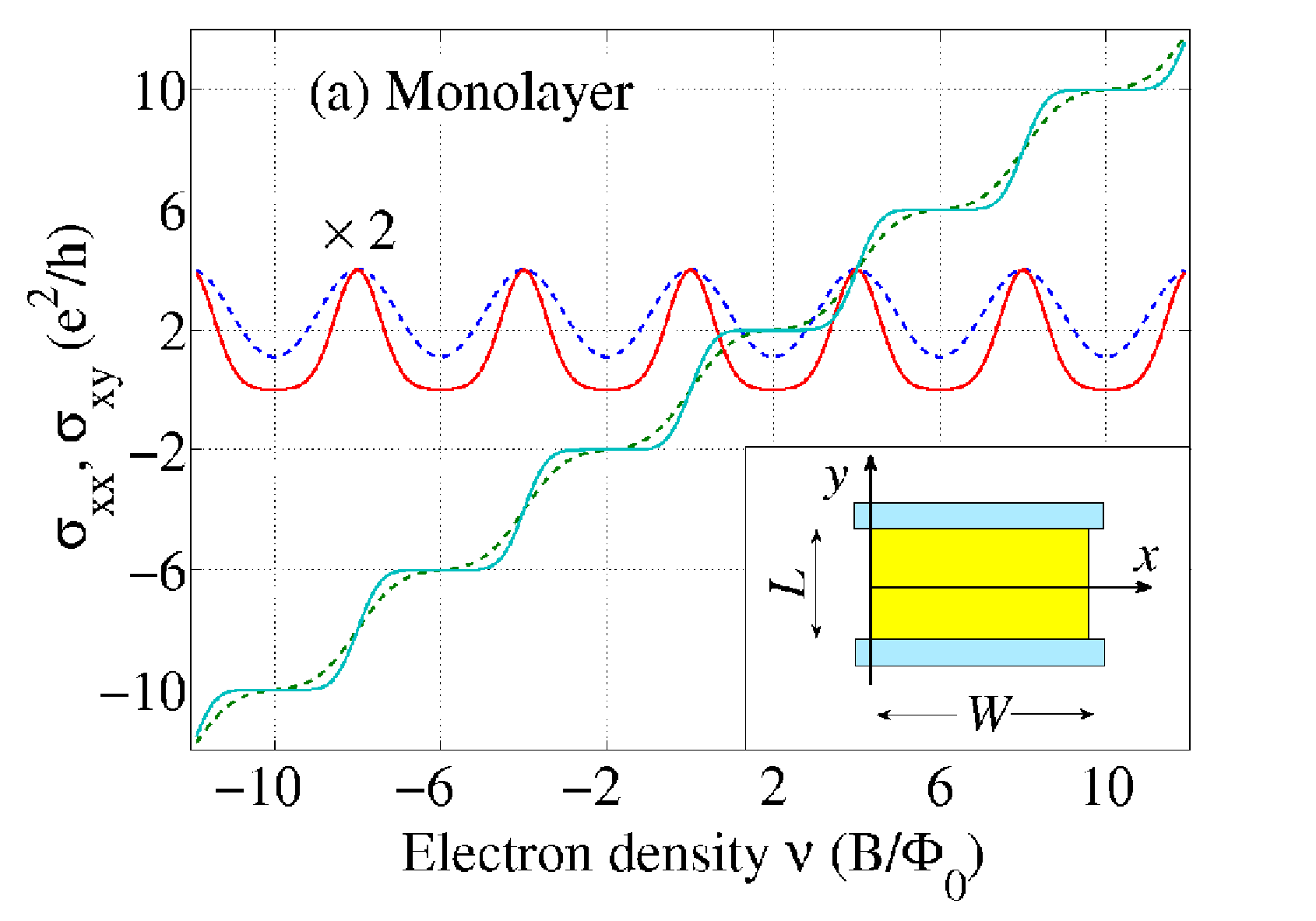}
\includegraphics[width=3.4in]{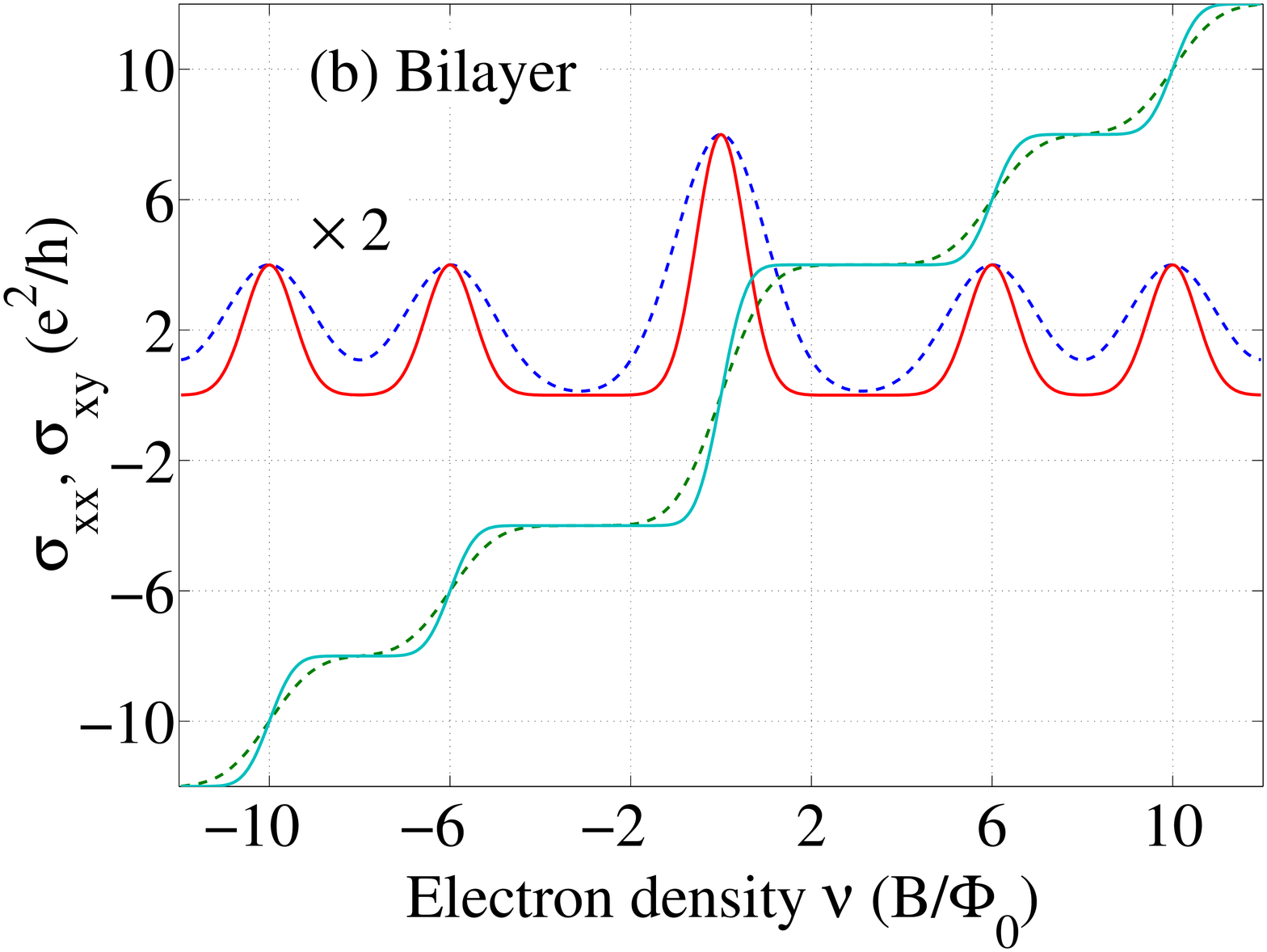}
\vspace{-0.1cm}
\caption[]{Longitudinal and Hall conductivity for (a) graphene monolayer and (b) graphene bilayer, obtained from the semicircle model, Eqs.(\ref{eq:sigma}),(\ref{eq:semicircle}),(\ref{eq:sxx_n}), for two values of the Landau level width parameter $\lambda=1.7$ (solid lines), $\lambda= 0.5$ (dashed lines). 
{\it Inset:} Schematic of a conducting sample of dimensions $L$ and $W$, of a rectangular shape, with source and drain at opposite sides.
}
\label{fig2}
\end{figure}

\section{The semicircle model}

The effective medium approach~\cite{semicircle} provides a convenient framework for
understanding the density dependence of $\sigma_{xx}$ and $\sigma_{xy}$ in QHE systems. It predicts a semicircle relation between $\sigma_{xx}$ and $\sigma_{xy}$, derived from a two-phase model, in which the system at the QHE transition is treated as a mixture of incompressible puddles with local Hall conductivities $\sigma_{xy}'$, $\sigma_{xy}''$, given by the quantized values at the neighboring QHE plateaus. 

The semicircle relation is a statement about properties of the macroscopic conductivity tensor on the length scales much greater than the size of individual puddles.
Although the validity of this relation, strictly speaking, is limited to the regime dominated by large-scale fluctuations of electron density, it is empirically known to provide a good description of the integer QHE plateau transitions observed in various semiconducting systems~\cite{Murzin02}. Recently, the semicircle relation was employed to describe transport coefficients in graphene~\cite{Abanin07a,Abanin07b,Burgess07}.

We stress that, 
while the semicircle model
is realistic, and also quite convenient to use, its specifics are almost certainly not essential for our conclusions. 
We believe that slight departure in behavior of transport coefficients from the semicircle model will have little effect on the properties of conductance.

Prior to turning to the semicircle model, we recall that
the conductivity $\sigma_{xy}$ in graphene monolayer exhibits steps of size $4e^2/h$ between adjacent integer quantum Hall plateaus, where the factor of four describes combined spin and valley degeneracy of Landau levels. The incompressible densities corresponding to the QHE plateaus in graphene monolayer are~\cite{Novoselov05,Zhang05} 
\be\label{eq:QHE1}
\nu_n=4(n+1/2)|B|/\Phi_0, \quad n=0,\pm 1, \pm 2...,
\ee
where $|B|/\Phi_0$ is electron density for a single Landau level.
In graphene bilayer QHE, due to accidental degeneracy of the Landau level positioned at the neutrality point $\nu=0$, there is an $8e^2/h$ Hall conductance step between the plateaus with $\sigma_{xy}=\pm 4e^2/h$, whereas other conductance steps are of normal $4e^2/h$ size. Accordingly,
in the bilayer the incompressible QHE densities are~\cite{Novoselov06} 
\be\label{eq:QHE2}
\nu_n=4n|B|/\Phi_0, \quad n=\pm 1, \pm 2...
\ee
In both cases (\ref{eq:QHE1}) and (\ref{eq:QHE2}) the density values $\nu_n$ are arranged symmetrically around the neutrality point. 
Hall conductivity on the plateaus takes the values $\sigma_{xy,n}^{(0)}=\nu_n e^2/h$, where $\nu_n$ are densities (\ref{eq:QHE1}) and (\ref{eq:QHE2}). 

In the semicircle model~\cite{semicircle}, the contributions of each Landau level to the longitudinal and Hall conductivities $\delta_n\sigma_{xx}(\nu)$, $\delta_n\sigma_{xy}(\nu)$ are related by the semicircle law: 
\be\label{eq:semicircle}
\delta_n\sigma_{xx}^2+(\delta_n\sigma_{xy}-\sigma_{xy,n}^{(0)})(\delta_n\sigma_{xy}-\sigma_{xy,n'}^{(0)})=0,
\ee
where $\sigma_{xy,n}^{(0)}$ and $\sigma_{xy,n'}^{(0)}$ are the quantized Hall conductivities on adjacent plateaus. Here $n$ and $n'$ are neighboring integers in the sequence $...-2,-1,0,1,2...$ for the monolayer, and $...-2,-1,1,2...$ for the bilayer: $n'=n+1$ except the double-degenerate $\nu=0$ Landau level for the bilayer, in which case $n=-1$, $n'=1$.

The longitudinal conductivity $\delta_n\sigma_{xx}(\nu)$ exhibits a peak centered at the Landau level. We model it by a gaussian 
\be\label{eq:sxx_n}
\delta_n\sigma_{xx}(\nu)=\frac 12 C_n e^{-\lambda\lp \nu-\frac12(\nu_n+\nu_{n'})\rp^2}
\ee
where the parameter $\lambda$ describes  broadening of the Landau level (large values of $\lambda$ correspond to a narrow Landau level). 

In the semicircular model, the peak value of $\delta_n\sigma_{xx}$ must equal to $\frac12(\sigma_{xy,n'}^{(0)}-\sigma_{xy,n}^{(0)})$. This is ensured by the prefactor $C_n$ value in Eq.(\ref{eq:sxx_n}) chosen to coincide with the $n$th Landau level degeneracy. For graphene monolayer we set $C_n=4$ (spin and valley degeneracy) for all $n$, while for the bilayer $C_n=8$ for $n=-1$ (spin, valley and accidental degeneracy) and $C_n=4$ for all other $n$'s.

The total conductivity tensor is given by the sum of the contributions of all Landau levels, 
\be\label{eq:sigma}
\sigma_{xx}(\nu)=\sum_{n} \delta_n\sigma_{xx}(\nu),\quad \sigma_{xy}(\nu)=\sum_{n} \delta_n\sigma_{xy}(\nu).
\ee
For simplicity, here we choose the same value of the parameter $\lambda$ for all Landau levels.
The resulting conductivity density dependence is illustrated in Fig.\ref{fig2}a,b. 

We point out that an interesting prediction can be drawn, specific to the graphene zeroth Landau level ($\nu=0$), by combining the semicircle model (\ref{eq:semicircle}) with Eq.(\ref{g_square}). In a square sample with a negligible overlap between Landau levels, the two-terminal conductance (\ref{g_square}) would be  completely density-independent across the zeroth Landau level, because in this case
$\sqrt{\sigma_{xx}^2+\sigma_{xy}^2}$ would equal $2\frac{e^2}{h}$ for the monolayer and $4\frac{e^2}{h}$ for the bilayer. This happens because the density dependence of transport coefficients, 
the peak in $\sigma_{xx}$ and the step in $\sigma_{xy}$, centered at $\nu=0$, cancel each other in the expression (\ref{g_square}). For weakly overlapping Landau levels, the contributions of the levels adjacent to $\nu=0$ would lead to slight deviations from a flat plateau.

The conductance measured in graphene indeed often exhibits a plateau across the entire $\nu=0$ region. Two examples of such behavior in recent literature are Ref.\cite{Heersche07}, Fig.\,1d, and Ref.\cite{Williams07}, Fig.\,3B. In both cases, the measured conductance is nearly flat in a wide density interval centered at $\nu=0$, with a small peak in the middle. Interestingly, the sample geometry in both cases was quite different from a square. In Ref.\cite{Heersche07} the AFM image indicated that the sample was approximately rectangular with contacts at opposite sides, similar to the schematic in Fig.\ref{fig2} inset. However, its width was quite large, $W\approx 5L$, in which case a fairly large peak at $\nu=0$ is to be expected (see below). To understand the discrepancy, it would be useful to known how spatially uniform the conduction was, in particular near contacts (poor contact along part of the sample edge could reduce the effective sample width, bringing 
 $W$ closer to $L$). As for the device in Ref.\cite{Williams07}, its contact geometry was quite far from rectangular, yet giving rise to a fairly good plateau around $\nu=0$. 

Such a behavior can be understood in the context of universality of the conductance problem that results from its conformal invariance. As we discuss in Appendix, for an arbitrary geometry of a conductor with two contacts of arbitrary sizes the conductance is identical to that of an equivalent rectangle. In other words, all possible conduction problems are classified by the values of a single parameter, the aspect ratio $L/W$ of an equivalent rectangle. Then, as long as $L/W\approx 1$, the conductance would behave in the same way as for a square shape, even if the actual geometry is very different from a square.


\begin{figure}
\includegraphics[width=3.4in]{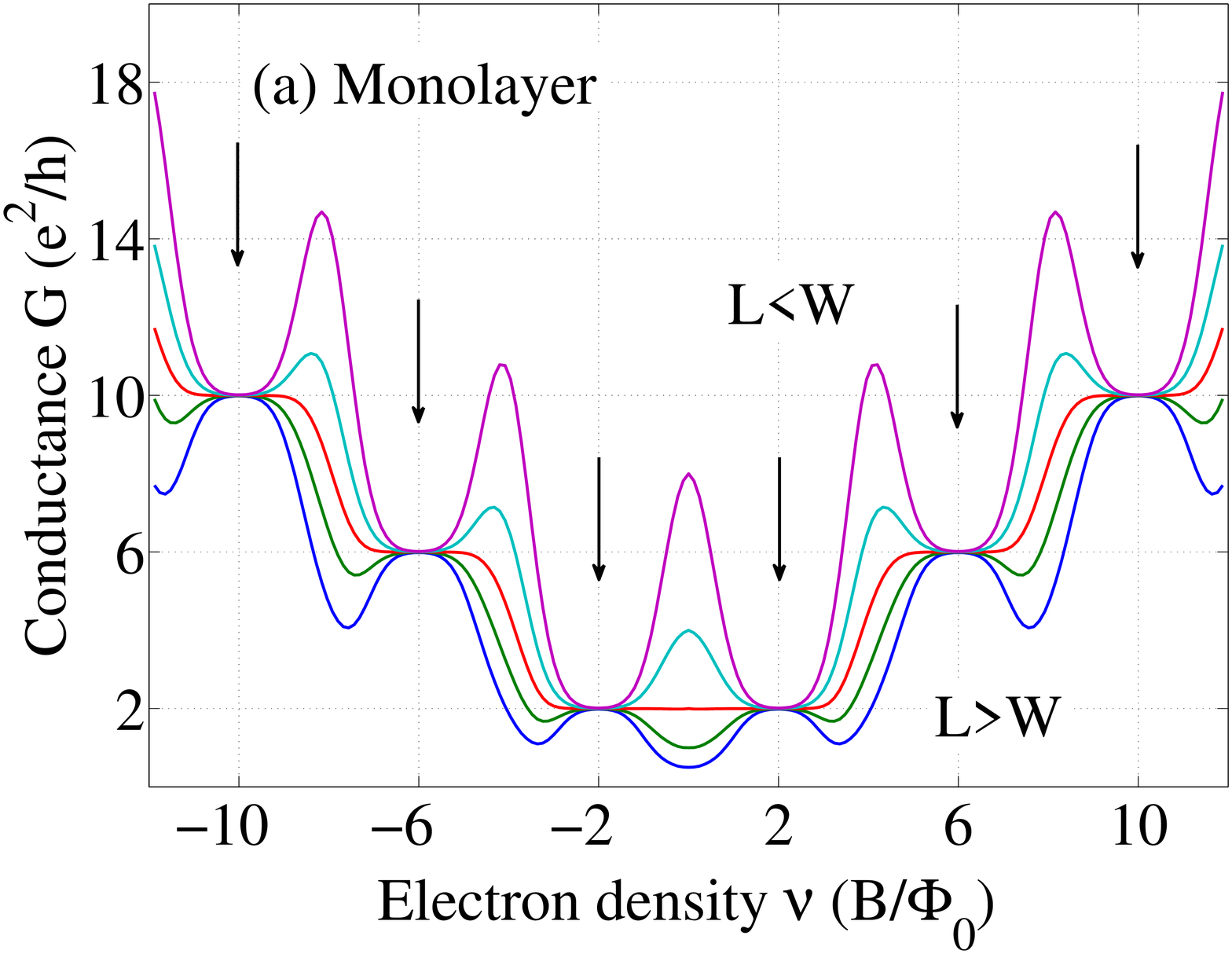}
\includegraphics[width=3.4in]{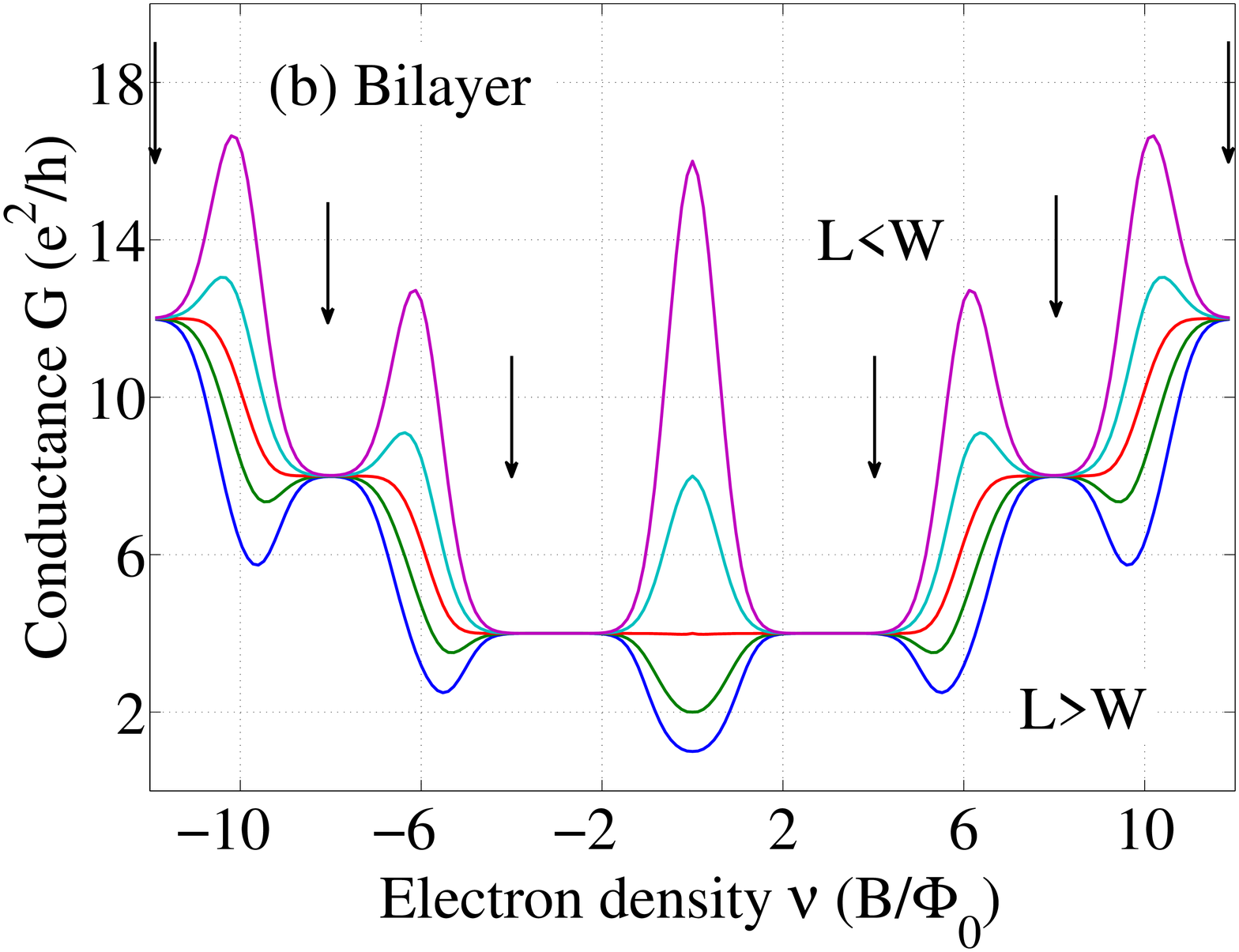}
\vspace{-0.1cm}
\caption[]{Two-terminal conductance (\ref{eq:conductance}) of a rectangular graphene sample: (a) monolayer, (b) bilayer, for  aspect ratios $L/W=0.25,0.5,1,2,4$ (top to bottom) with the Landau level width parameter $\lambda=1.7$ (corresponding to solid lines in Fig.\ref{fig2}). Arrows mark the incompressible densities (\ref{eq:QHE1}), (\ref{eq:QHE2}). Note the plateau at $\nu=0$ for the square case, $L=W$ (red curve), which is in agreement with the behavior predicted by Eq.(\ref{g_square}).
}
\label{fig4}
\end{figure}

\section{Conformal mapping approach}

Now we proceed to describe the distribution of current and electric field in a rectangular sample with an arbitrary aspect ratio $L/W$ and spatially uniform conductivities $\sigma_{xx}$, $\sigma_{xy}$ (see Fig.\ref{fig2} inset). This problem, with zero boundary condition for normal current component at the sample boundary and for tangential field component at the sample/contact interface, has been treated in Refs.~\cite{Lippmann58,Girvin81} using conformal mapping technique.

In Ref.~\cite{Lippmann58} the Schwartz-Christoffel mapping was employed, leading to the following 
integral representation for the conductance:
\be\label{integral_G}
G=\sqrt{\sigma_{xx}^2+\sigma_{xy}^2}\, \frac{I(1,-1)}{I(1/k,1)}, 
\ee  
where 
\be\label{I()}
I(u,v)=\int_{v}^{u} \frac{d\xi}{|(\xi-1)(1/k+\xi)|^{\delta_+}|(\xi+1)(1/k-\xi)|^{\delta_-}}
.
\ee
Here $\delta_\pm=1/2\pm \theta/\pi$, and $\theta={\tan}^{-1} (\sigma_{yx}/\sigma_{xx})$ is the Hall angle. The parameter $k$ above is the elliptic modulus, $0<k<1$, related to the sample aspect ratio via
\be\label{K(k)}
\frac{L}{W}=\frac{K(k)}{2K(k')}
, \quad k'=\sqrt{1-k^2}
,
\ee
where $K(k)$ is the complete elliptic integral of the first kind. 

In principle, Eqs.(\ref{integral_G}),(\ref{I()}),(\ref{K(k)}) give a complete solution of our problem. However, the integrals in Eq.(\ref{integral_G}) contain power-law singularities with the exponents $\delta_\pm=1/2\pm \theta/\pi$ that can approach unity for $\theta\approx \pm\pi/2$. This makes the integrals in Eq.\eqref{I()} difficult to evaluate numerically, especially for large Hall angles, 
when the singularities are the strongest and the integrals are converging fairly slowly.

Instead, we use a different approach, developed in Ref.~\cite{Girvin81}, which provides an expression of the electric field and current in the sample in terms of the exponentials of infinite but rapidly convergent sums. This method is more convenient for our purposes, because it allows to choose the integration contour for numerical evaluation of the conductance so that it bypasses the singularities.

In Ref.~\cite{Girvin81}, the electric field components $E_x$ and $E_y$ at a point $z=x+iy$ are obtained as the real and imaginary parts of a suitable analytic function. The latter function is found to be of the form
\be\label{eq:el_field}
E_y+iE_x=-e^{f(z)}, 
\ee
where 
\be\label{eq:k_Z}
f(z)=i\theta-\sum_{n>0\,{\rm (odd)}} \frac{4\theta}{n\pi} \, \frac{\sinh(n\pi i z/W)}{\cosh(n\pi L/2W)}  . 
\ee
Current distribution can be obtained by combining (\ref{eq:el_field}) with the relation
$j_x+ij_y=(\sigma_{xx}+i\sigma_{yx})(E_x+iE_y)$.

\begin{figure}
\includegraphics[width=3.4in]{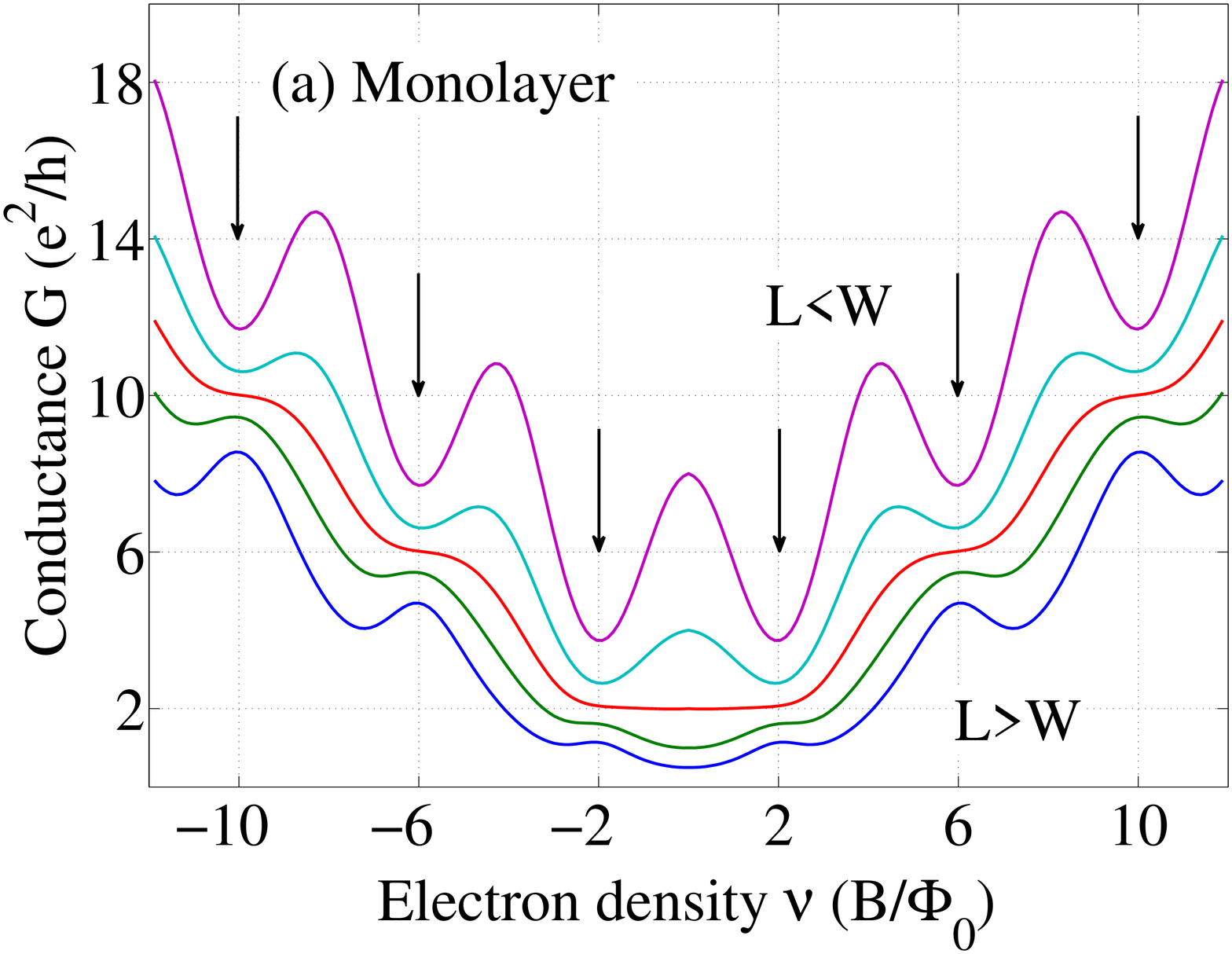}
\includegraphics[width=3.4in]{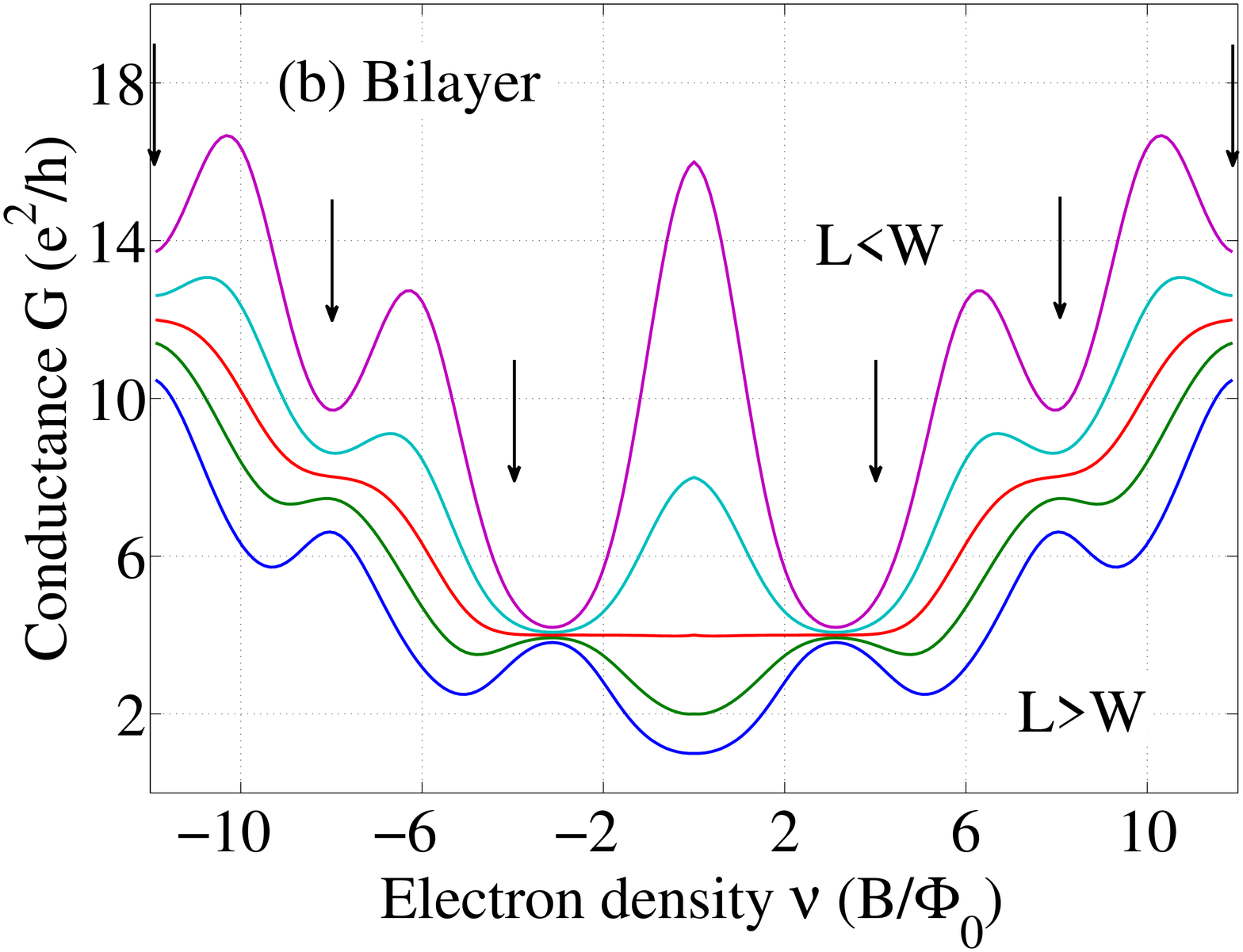}
\vspace{-0.1cm}
\caption[]{
Same as in Fig.\ref{fig4} for broader Landau levels, described by (\ref{eq:sxx_n}) with $\lambda=0.5$ (dashed lines in Fig.\ref{fig2}); the sample aspect ratios are $L/W=0.25,0.5,1,2,4$ (top to bottom).
Note that the qualitative features, such as the positions of the conductance minima at the QHE plateau centers for $L<W$ (maxima for $L>W$), as well as the conductance values at these densities, are similar to those seen in Fig.\ref{fig4} despite increased Landau level broadening. Note also the relative size of the $\nu=0$ peak in the monolayer and bilayer cases, compared to the size of neighboring peaks at other compressible densities, which is also insensitive to the Landau level broadening. 
}
\label{fig6}
\end{figure}

Once the current distribution is found, it can be used to obtain the two-terminal conductance
\be\label{eq:conductance}
G=I/V, 
\ee
where $I$ is the total current and $V$ is the source-drain bias voltage. 
To evaluate the net current $I$ by integrating current density, 
one has to choose 
a cross-section through the sample that does not pass through its corners, where the function $f(z)$ has singularities. It is particularly convenient to 
perform this integration along a straight line that cuts through the middle of the rectangle at $y=0$ (along the $x$-axis in Fig.\ref{fig2} inset):
\be\label{eq:total_current}
I=\int_0^W j_y(x,0) dx
. 
\ee
Since $j_y=\sigma_{xx} E_y+
\sigma_{yx} E_x$, the current is given by 
\be\label{eq:total_current2}
I=\int_0^W \sigma_{xx}(E_y(x,0)+\tan\theta E_x(x,0)) dx.  
\ee
For similar reasons, we calculate the voltage drop between the upper and 
lower contacts ($y=\pm L/2$) as an integral of the electric 
field along a straight line connecting the points 
$(W/2,-L/2)$ and $(W/2,L/2)$ of the contacts:
\be\label{eq:voltage}
V=-\int_{-L/2}^{L/2} E_y(W/2,y) dy. 
\ee
Evaluating the integrals (\ref{eq:total_current2}) and (\ref{eq:voltage}) numerically, we obtain the conductance (\ref{eq:conductance}) as a function of transport coefficients, which defines its dependence on electron density. The results for different aspect ratios $L/W$ 
for the monolayer and the bilayer case
are displayed in Fig.\ref{fig4} and Fig.\ref{fig6} for two different values of the Landau level width parameters in (\ref{eq:sxx_n}), $\lambda=1.7,\,0.5$.

As evident from Fig.\ref{fig4}, the conductance behavior depends strongly on the sample aspect ratio $L/W$. For a square sample, $L=W$, the conductance is a monotonic function of density at positive and negative $\nu$, rising in steps associated with QHE plateaus. In this case, the behavior of $G$ is found to be completely flat near $\nu=0$ in agreement with the above discussion based on Eq.(\ref{g_square}).

For $L\ne W$, the conductance turns into a nonmonotonic function of density, with the QHE plateaus being less pronounced than for $L=W$. For $L<W$ the conductance has minima near QHE plateau centers, Eq.(\ref{eq:QHE1}) 
for the monolayer, and Eq.(\ref{eq:QHE2})
for the bilayer case (marked by arrows in Figs.\ref{fig4},\ref{fig6}), while for $L >W$ the conductance has maxima at these densities. Overall, the conductance behavior for $L<W$ is ``inverted'' compared to that for $L>W$, as expected from the duality relation (\ref{duality}) which implies 
\[
G_{L<W}>G_{L=W}>G_{L>W}
\]
for all $\nu$. Recently, the conductance as a function of carrier density was studied for samples of several aspect ratios~\cite{Williams08}, with the results being in qualitative agreement with ours.

With these results at hand we are now in position to ask which differences between the monolayer and bilayer systems are robust with respect to Landau levels' broadening and variation in sample geometry. First, we note that the centers of the QHE plateaus (arrows in Figs.\ref{fig4},\ref{fig6}) remain close to the positions of the minima in $G$ for $L<W$, and its maxima for $L>W$. The corresponding incompressible values of the density, Eqs.(\ref{eq:QHE1}),(\ref{eq:QHE2}), 
which are marked by arrows in Figs.\ref{fig4},\,\ref{fig6},
are equally spaced in the monolayer case, but are {\it not equally spaced} in the bilayer case due to the eight-fold Landau level degeneracy at $\nu=0$. It can be seen by comparing Fig.\ref{fig4} and Fig.\ref{fig6} that this difference between the monolayer and bilayer systems is not masked by Landau levels' broadening or by the aspect ratio variation.

Next, the conductance values at the minima for $L<W$ and at the maxima for $L>W$ remain close to the associated QHE values $G\approx \nu_n e^2/h$, which are different in the monolayer and bilayer cases. This difference is clear in Figs.\ref{fig4},\ref{fig6} even for large aspect ratios and broadened Landau levels. In practice, however, these values may change as a result of added contact resistance. If this is the case, the relative positions of the plateaus in density, inferred from the arrangement of maxima and minima of $G$, are more robust than the conductance values at these densities. 

Finally, we note the difference in the size of the peak at $\nu=0$  as compared to the sizes of neighboring peaks at other compressible densities, found for the monolayer and bilayer at $L<W$. (The same is true for the dip at $\nu=0$ for $L>W$.) In the monolayer case, the size of the $\nu=0$ peak/dip is comparable to the sizes of neighboring peaks/dips, whereas in the bilayer case the $\nu=0$ peak/dip is almost two times larger than its neighbors. This difference reflects the higher density of states in the $\nu=0$ Landau level of a bilayer. 

It is also interesting to note that the heights of all peaks and dips in $G$ are completely independent of the Landau level broadening, as can be seen by comparing the curves in Fig.\ref{fig4} and Fig.\ref{fig6}. 
This behavior is specific to the semicircle model, in which the peak values of $\sigma_{xx}$ and corresponding values of $\sigma_{xy}$ are universal. Indeed, as evident from Fig.\ref{fig2}, peaks of $\sigma_{xx}$ line up with steps in $\sigma_{xy}$, and the values of $\sigma_{xy}$ in the middle of each step are independent of the Landau level broadening. 
As a result, although broadened Landau levels change the overall behavior of the conductance $G$, the peak values remain intact.

\section{Conductance fluctuations in small samples}

So far we have described the system by a spatially uniform conductivity, assuming this effective medium model to be valid at all filling factors. This assumption, adequate for large samples, breaks down for smaller samples when the sample dimensions $L$ or $W$ become comparable to the typical charge inhomogeneity length 
scale $\xi$. Conductance of such small samples at QHE transitions can exhibit strong sample-to-sample fluctuations and significant deviations from the prediction of the semicircle model. 

The effect of charge inhomogeneity is expected to be especially strong near the charge-neutrality point (CNP), 
where the density of free carriers is low and screening of the disorder potential is poor, for both the monolayer and bilayer graphene. For simplicity, below we shall focus on the monolayer graphene in the vicinity of the CNP, $\nu\approx 0$.  To qualitatively understand the effect of charge inhomogeneities, we employ the two-phase model~\cite{semicircle}
treating
the sample as a mixture of incompressible puddles of types $n$ and $p$ of typical size $\xi$. The filling factors in the puddles are $\nu=\pm 2$, 
corresponding to the QHE plateaus adjacent to CNP. 

\begin{figure}
\includegraphics[width=3in]{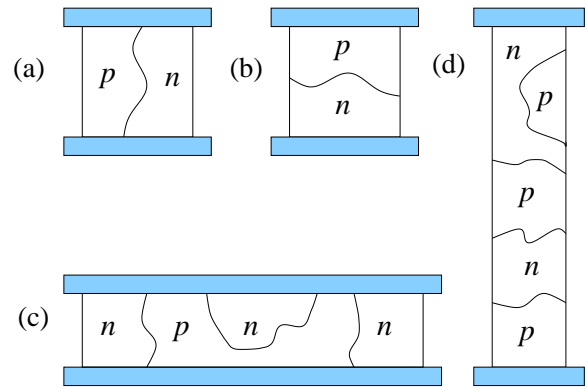}
\vspace{-0.1cm}
\caption[]{Examples of possible {\it p}-type and {\it n}-type puddle configurations near graphene charge neutrality point for mesoscopic samples with at least one of the dimensions comparable to the typical puddle size $\xi$. In (a) and (b) two possible configurations are shown for a square sample with $L,\,W\approx \xi$. The corresponding conductance values, $G=4e^2/h$ for (a) and $G=e^2/h$ for (b), are different from the value $G^*=2e^2/h$ predicted by the effective medium (semicircle) model.  
Puddle arrangement in a short and wide sample ($L\lesssim \xi\ll W$) and in a narrow and long sample ($W\lesssim \xi\ll L$) is illustrated in (c) and (d). }
\label{fig5}
\end{figure}

To illustrate the effect of strong fluctuations near the CNP, we first inspect the case of a small square sample, with 
both dimensions $L$ and $W$ comparable to $\xi$. Suppose that the carriers in such a sample form just two puddles, of type $p$ and $n$ ($\nu=\pm 2$). There are two topologically distinct puddle configurations giving the net conductance different from the value $G^*=2e^2/h$ expected from the semicircle relation: (i) the {\it p-n} boundary connecting the source and the drain  (see Fig.\ref{fig5}a) (ii) the {\it p-n} boundary connecting the opposite free edges of the sample (see Fig.\ref{fig5}b). 

In the case (i) the {\it p} and {\it n} regions are connected in parallel, and thus
the net conductance is the sum of the conductances $G_0=2e^2/h$,
giving $G=2G^\ast=4e^2/h$. In the case (ii) the {\it p} and {\it n} regions are connected in series, with the net conductance equal $G=1/2G^\ast=e^2/h$ (see Ref.~\cite{Abanin07c}). Thus for mesoscopic square samples with $L,\,W \sim \xi$ the conductance strongly depends on the puddle configuration and can exhibit either a peak or a dip at $\nu=0$, 
whereas the semicircle model predicts a plateau.  

Now let us consider a sample which is short but wide: $L\lesssim \xi\ll W$ (see Fig.\ref{fig5}c). Such a sample consists of about $N\sim W/\xi$ 
alternating {\it p} and {\it n} regions of width $\sim \xi$. The total conductance is given by the sum of the conductances of the individual regions, the conductance of each region being $G_0=2 e^2/h$. 
This yields the net conductance of $G\sim N G_0=(2W/\xi) \frac{e^2}{h}$. This is about $L/\xi$  smaller than the value $G^*=(2W/L)\frac{ e^2}{h}$ predicted by the semicircle relation for the effective medium model. 

Similarly, for the case of a narrow but long sample, $W\lesssim \xi\ll L$ (see Fig.\ref{fig5}d) the conductance is given by the series conductance of $N\sim W/\xi$ puddles: $G\sim (2 \xi/L) \frac{e^2}{h}$. This is $\xi/W$ times larger than than the prediction of the effective medium model. These estimates show that near CNP the semicircle model overestimates the conductance of very short samples, and underestimates the conductance of very narrow samples, when 
at least one of the sample dimensions is comparable to the puddle size.

\section{Conclusions}

The above results for the two-terminal conductance of rectangular samples, and in particular, the conductance dependence on the sample aspect ratio, can serve as a benchmark for understanding properties of graphene samples. 
We found that $G$ exhibits peaks (dips) at the compressible densities for $L<W$ ($L>W$), which completely disappear at $L=W$. 
We could identify several specific features that may help to distinguish between transport in graphene monolayer and bilayer. Those include positions of the incompressible densities, inferred from minima (maxima) in $G$ at $L<W$ ($L>W$), the values of $G$ at these densities, and the relative size of the central peak (dip) in $G$ as compared to the neighboring peaks. These features, which 
are shown to be insensitive to the sample aspect ratio and to Landau levels' broadening within our model, can be used for sample diagnostic in transport measurements.

We thank 
Charles Marcus, Leo Di Carlo, James Williams, Pablo Jarillo-Herrero, and Alberto Morpurgo for useful discussions. This work was supported by the grant NSF-NIRT DMR-0304019.

\appendix

\section{Appendix: Rectangle as the mother of all shapes; Conformal invariance and universality}

There is a profound relation between the two-terminal conductance and conformal invariance of the 2d transport problem. It arises because for 2d conductors of arbitrary shapes with spatially uniform $\sigma_{xx}$ and $\sigma_{xy}$ 
the conductance is invariant under conformal transformations, and because all single-connected domains in the plane can be conformally mapped to one another. As a result, a conducting domain of any shape has the same conductance as a particular domain of some simple shape with an appropriate arrangement of contacts. The simple shape can be chosen in a number of ways, in particular it can be chosen to be a rectangle. 
We show in this section that for a conductor of any shape with any configuration of two contacts {\it the conductance is equal to that of a rectangle} with some aspect ratio $L/W$.

Because the correspondence between conductors of arbitrary shapes and equivalent rectangles is purely geometric (it is defined by a conformal mapping), the aspect ratio $L/W$ of an equivalent rectangle depends on the sample shape but does not depend on the values of transport coefficients $\sigma_{xx}$ and $\sigma_{xy}$. As a result, one can use the rectangle problem with a fixed aspect ratio to describe conductance as a function of the carrier density, via the $\sigma_{xx}$ and $\sigma_{xy}$ density dependence.

To formulate the constraints due to conformal invariance, 
we recall that conformal mappings in 2d are realized as analytic functions of the complex variable $z=x+iy$. Thus we consider mappings between two complex planes $z$ and $w$ defined by analytic functions $w=f(z)$, which map the sample domain (hereafter denoted ${\cal D}$) in the plane $z$ onto a domain ${\cal D}'$ in the plane $w$. 

The relations \eqref{div-curl} as well as the boundary conditions $E_\parallel=0$ on the contacts and $j_\perp=0$ on the open boundary are invariant under such conformal mappings. The easiest way to verify this is to note that the current continuity condition $\nabla\cdot \vec j=0$ can be solved by $\vec j=\nabla\times (\vec z \phi(x,y))$, where $\vec z$ is the unit vector normal to the plane. Then the relation $\nabla\times \vec E=0$ combined with $j_x+ij_y=(\sigma_{xx}+i\sigma_{yx})(E_x+iE_y)$ means that the function $\phi$ is harmonic, i.e. it satisfies the Laplace's equation 
\be\label{eq:Laplace's}
\lp \partial_x^2+\partial_y^2\rp \phi(x,y)=0
. 
\ee
%
Because a harmonic function remains harmonic under a conformal mapping, and because the angles between the gradient $\nabla \phi(z)$ and the boundary of ${\cal D}$ are the same as the angles between  $\nabla \phi(w)$ and the boundary of ${\cal D}'$ at corresponding points, Eqs.\eqref{div-curl} as well the relations $E_\parallel=0$ and $j_\perp=0$ are indeed conformally invariant.

\begin{figure}
\includegraphics[width=3in]{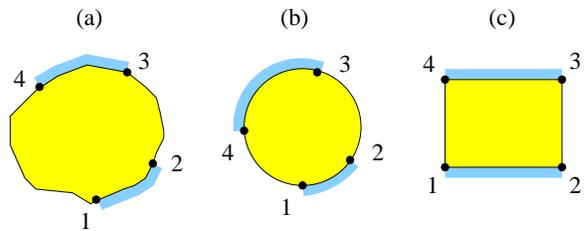}
\vspace{-0.1cm}
\caption[]{ Conformal invariance illustrated by several conducting domains with contacts. If two domains can be conformally mapped on each other so that the contact regions are mapped on the corresponding contact regions, their conductances are the same. The Riemann mapping theorem guarantees existence of a mapping between an arbitrary domain (a) and a unit disk (b) such that three points  on the boundary of (a), marked 1, 2 and 3, are mapped on any three points (1, 2 and 3) on the circle (b). 
The position of the fourth point, which is not specified, defines a one-parameter family of possible conduction problems. All such problems can be parameterized by rectangles (c) with different aspect ratios.
}
\label{fig7}
\end{figure}

This simple mathematical fact can serve as a platform for the following interesting observation. Suppose an analytic function $w=f(z)$  maps the conducting domain  ${\cal D}$ in the plane $z$ to a domain ${\cal D}'$ in the plane $w$ so that a pair of contacts to ${\cal D}$ is mapped on a pair of contacts to ${\cal D}'$. More precisely, let 
four points $1,2,3,4$ on the boundary of ${\cal D}$ be mapped onto four points $1',2',3',4'$ on the boundary of ${\cal D}'$. Let the arcs $1-2$, $3-4$ and $1'-2'$, $3'-4'$ be ohmic contacts for the problems in the $z$ and $w$ planes, respectively (see Fig.\ref{fig7}). Then the two-terminal conductance of ${\cal D}$ is exactly the same as that of ${\cal D}'$, i.e the conductance is invariant under all conformal mappings that map to one another the corresponding conducting domains and contacts.

On the other hand, as is well known from the theory of complex variables
(the Riemann mapping theorem), 
any two single-connected domains ${\cal D}$ and ${\cal D}'$ can be mapped onto each other. This mapping can be fixed so that any three points on the boundary of ${\cal D}$ are mapped to any three points on the boundary of ${\cal D}'$
(under these conditions the mapping is unique).
This allows to reduce the conductance problem of an arbitrary domain ${\cal D}$ to that of some simple domain, e.g. a circular disk $|w|\le 1$.

Thus the only reason the conductors of different shapes do not all have the same conductance is the additional freedom in choosing the contacts, defined by the points $1,2,3,4$. 
Furthermore, because any conduction problem is equivalent, via a conformal mapping, to a disk with contacts defined by four points on the boundary $z_1,...,z_4=e^{i\theta_1},...,e^{i\theta_4}$, and because three of those points can be fixed by the Riemann mapping theorem, the only freedom left is in the position of the fourth point. Therefore, all conductance problems form {\it a one-parameter family}. For the points $z_1,...,z_4$ this parameter can be expressed, e.g., as the so-called cross ratio
\be\label{eq:cross-ratio}
\Delta_{1234}=\frac{(z_1-z_4)(z_3-z_2)}{(z_1-z_2)(z_3-z_4)}
\ee
which takes real values for any four points that lie on a circle. Applied to the problem of a rectangle, this procedure would give a one-to-one relation between the parameters $L/W$ and $\Delta_{1234}$, proving that indeed any conductance problem is isomorphic to that of a rectangle.

\begin{figure}
\includegraphics[width=3in]{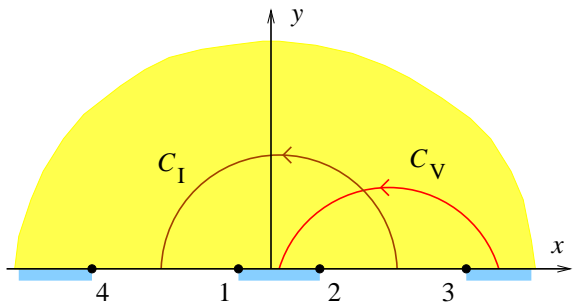}
\vspace{-0.1cm}
\caption[]{Complex halfplane and conduction problem in it, which can be  mapped on that in a rectangle using the Schwartz-Christoffel mapping, Eq.\eqref{eq:Schwartz-Christoffel}. The aspect ratio $L/W$ of the equivalent rectangle depends on the relative size of the contacts, shown in blue, and the distance between them. The end points of the contacts are $\xi_{1,2}=\mp 1$, $\xi_{3,4}=\pm 1/k$.
}
\label{fig8}
\end{figure}

To obtain such a relation between $L/W$ and $\Delta_{1234}$, let us consider a complex halfplane ${\rm Im}\,\xi>0$ with four points on the real axis 
\be\label{eq:xi1234}
 \xi_{1,2}=\mp 1 
,\quad \xi_{3,4}=\pm 1/k
, 
\ee
with the parameter $0<k<1$ (see Fig.\ref{fig8}). 
Under a fractional-linear mapping that maps the unit disk to the halfplane, the points $z_1,...,z_4$ can be mapped onto corresponding points \eqref{eq:xi1234} if and only if the cross ratios are the same:
\be\label{eq:Delta-k}
\Delta_{1234}=(1-k^2)/4k
\ee
(the cross ratio is an invariant of fractional linear mappings). On the other hand, using the Schwartz-Christoffel formula, the halfplane ${\rm Im}\,\xi>0$ can be mapped on a rectangular domain by the function 
%
\be\label{eq:Schwartz-Christoffel}
z=F(\xi,k)\equiv\int_{0}^\xi\frac{d\xi'}{\sqrt{(1-\xi^{\prime 2})(1-k^2\xi^{\prime 2})}}
,
\ee
where $F(\xi,k)$ is the elliptic integral of the first kind. This function maps $\xi_{1,2}=\mp 1$ to $z_{1,2}=\mp K(k)$, where
\be
 K(k)=\int_0^1\frac{d\xi}{\sqrt{(1-\xi^2)(1-k^2\xi^2)}} 
\ee
is the complete elliptic integral of the first kind. The points $\xi_{3,4}=\pm 1/k$ are mapped to $z_{3,4}=\pm K(k)+iK(k')$, where $k'=\sqrt{1-k^2}$.
Thus the sides of the rectangle have lengths $W=2K(k)$ and $L=K(k')$, which gives the relation between $L/W$ and $k$ of the form \eqref{K(k)}. Combining this with \eqref{eq:Delta-k}, we can relate $L/W$ to the cross ratio $\Delta_{1234}$.

In the halfplane of Fig.\ref{fig8} the distribution of the electric field and current can be found by noting that the function $E(\xi)=E_y+iE_x$ is analytic at ${\rm Im}\,\xi> 0$ and at the boundary its argument takes fixed values between the points $\xi_1,...,\xi_4$: 
${\rm arg}\,E_{-1<\xi<1}=0$, ${\rm arg}\,E_{1<\xi<1/k}=\frac{\pi}2+\theta$,
${\rm arg}\,E_{-1/k<\xi<-1}=-\frac{\pi}2+\theta$, ${\rm arg}\,E_{|\xi|>1/k}=-\pi$.
These requirements are sufficient to reconstruct the function:
\be
E(\xi)=\frac{A}{(1-\xi)^{\delta_+}(1+\xi)^{\delta_-}(1-k\xi)^{\delta_-}(1+k\xi)^{\delta_+}}
,
\ee
where $\delta_\pm=\frac12\pm\frac{\theta}{\pi}$ with $\theta=\tan^{-1}\sigma_{yx}/\sigma_{xx}$ the Hall angle, as in \eqref{I()}, and the unknown prefactor $A$ depends on the source-drain voltage. Current distribution is then found as $J(\xi)=j_y+ij_x=(\sigma_{xx}+i\sigma_{xy})E(\xi)$. 
The total current and voltage can be expressed by integrating $J(\xi)$ over the contour $C_I$, and $E(\xi)$ over the contour $C_V$ (see Fig.\ref{fig8}). The ratio of these integrals gives the result \eqref{integral_G}.

To summarize our discussion, because the conductor of an arbitrary shape can be conformally mapped on a rectangle which has the same conductance, the rectangle problem is ``universal.'' For a sample of any shape with a pair of contacts of arbitrary size and form, an equivalent rectangle can be found such that it has the same conductance.
Crucially, the aspect ratio $L/W$ of this rectangle is independent of the values of transport coefficients, which means that its conductance will have the same density dependence as that of the physical sample.

We note that, although the results of this section apply to conductors of completely arbitrary shapes, and in that sense they are far reaching, there are several limitations. First, we have assumed that the electron system is spatially uniform and homogeneous, i.e. the transport coefficients $\sigma_{xx}$ and $\sigma_{xy}$ are position-independent. Our second assumption was that transport is fully described by the 2d current-field relation $\vec j=\hat\sigma\vec E$, where $\hat\sigma$ is the conductivity tensor. In particular, the above model does not allow for the edge current states, which could, in principle, alter the shape dependence of the conductance. Still, since the bulk transport model agrees with the edge transport model in the limit of a large Hall angle, it can probably provide a good guidance even when the quantum Hall effect is fully developed.


\end{document}